\pdfoutput=1
\documentclass[aps,pra,a4paper,10pt,twocolumn,superscriptaddress]{revtex4-1}
\usepackage[american]{babel}
\usepackage[cp1252]{inputenc}
\usepackage[T1]{fontenc}
\usepackage{lmodern}

\usepackage[DIV=15]{typearea}

\usepackage{textcomp}
\usepackage{graphicx}

\newcommand{\TUKL}{Department of Physics and Research Center OPTIMAS, University of Kaiserslautern, Germany}
\newcommand{\IPM}{Fraunhofer Institute for Physical Measurement Techniques IPM, Freiburg, Germany}

\begin{document}

\title{Metamaterial-based gradient index lens with strong focusing in the THz frequency range}

\author{J.~Neu}
\altaffiliation{These authors contributed equally to this work.}
\affiliation{\TUKL}
\affiliation{\IPM}

\author{B.~Krolla}
\altaffiliation{These authors contributed equally to this work.}
\affiliation{\TUKL}
\affiliation{\IPM}

\author{O.~Paul}
\affiliation{\TUKL}

\author{B.~Reinhard}
\affiliation{\TUKL}

\author{R.~Beigang}
\affiliation{\TUKL}
\affiliation{\IPM}

\author{M.~Rahm}
\email[Electronic address: ]{\url{mrahm@physik.uni-kl.de}}
\affiliation{\TUKL}
\affiliation{\IPM}

\begin{abstract}
\noindent
The development of innovative terahertz (THz) imaging systems has recently moved in the focus of scientific efforts due to the ability to screen substances through textiles or plastics. The invention of THz imaging systems with high spatial resolution is of increasing interest for applications in the realms of quality control, spectroscopy in dusty environment and security inspections. One of the main restrictions of current THz imaging systems is the low spatial resolution which is limited by a lack of THz lenses with strong focusing capabilities. Here we present the design, fabrication and the measurement of the optical properties of spectrally broadband metamaterial-based gradient index (GRIN) lenses that allow one to focus THz radiation to a spot diameter smaller than the wavelength. Due to the subwavelength thickness and the high focusing strength the presented GRIN lenses are an important step towards compact THz imaging systems with strongly improved spatial resolution.
\end{abstract}

\maketitle

\noindent
THz physics and metamaterials have raised a great deal of scientific interest in the last few years. While recent advances in the THz technology were strongly driven by efforts to transfer the fundamental knowledge about THz radiation to industrial applications, the metamaterial science -- as the younger of both research fields -- was mainly dedicated to the investigation of the fundamental physical properties of metamaterials. This especially applies to metamaterials at higher frequencies where high absorptive loss  \cite{Dolling:2007} and the enormous demands on sophisticated three-dimensional nano-fabrication techniques \cite{Ergin:2010,Freymann:2010} obstructs the development of competitive and applicable metamaterial-based optics. In respect thereof, the limitations lead to the pursuit of many interesting basic research topics such as e.\ g.\ metamaterials with gain \cite{Xiao:2010,Lagarkov:2010,Fang:2010,Zheludev:2008} or the investigation of quantum effects in metamaterials \cite{Plumridge:2008,Kaestel:2005,Liu:2009}. In contrast, metamaterials at microwave frequencies are advantageous in terms of low optical loss and ease of fabrication. In consequence, microwave metamaterials have been developed to a maturity level that enables the integration into applied systems \cite{Rayspan:2010}.

Another promising frequency band for the application of metamaterials is the range from 0.1 to 10\,THz, the so-called THz range. Whereas THz physics was mostly concentrated on fundamental research about 7 years ago, the rapid development of the technology and the fabrication of improved THz emitters and detectors lead to the opening up of new markets and application fields in the last few years. The specific property of THz radiation to penetrate through most dielectrics as e.\ g.\ textiles, paper, concrete etc.\ has been exploited in many scenarios that are related to non-invasive quality control through packages \cite{Hoshina:2009,Rutz:2006,Herrmann:2002}, chemometrics of pharmaceutical substances \cite{Taday:2004}, terahertz imaging systems \cite{Mittleman:1996,Tonouchi:2007} and security inspections \cite{May:2007,Yamamoto:2004,Weg:2009,Kemp:2003}. The rapid expansion of industrial target applications is attended by increasing demands for high quality optical components for the THz technology. Moreover, many potential applications cannot yet be addressed by the THz technology since the required components are not available. However, such optics cannot be easily devised since the insufficient electromagnetic response of most dielectrics prevents the development of standard optics that are based on the same principles as conventional optics used in laser systems. In this context, metamaterials have already proven to offer a novel approach for modulating THz radiation \cite{Chen:2006,Chen:2008,Chen:2009,Paul:2009}, perfect absorbers with designable properties \cite{Landy:2008,Tao:2008}, frequency filters \cite{Paul:2009a} or wave plates \cite{Weis:2009}. Another application field where metamaterials can provide an important contribution is the THz imaging technology. One of the main restrictions of current compact state-of-the-art THz imaging systems is the low spatial resolution which is limited to about 1\,mm due to a lack of available THz lenses with strong focusing capabilities. It is obvious that the optical resolution of such imaging systems could be significantly improved if lenses with strong focusing capabilities and low aberrations were available.
\begin{figure*}
\centering
\includegraphics[scale = 0.7]{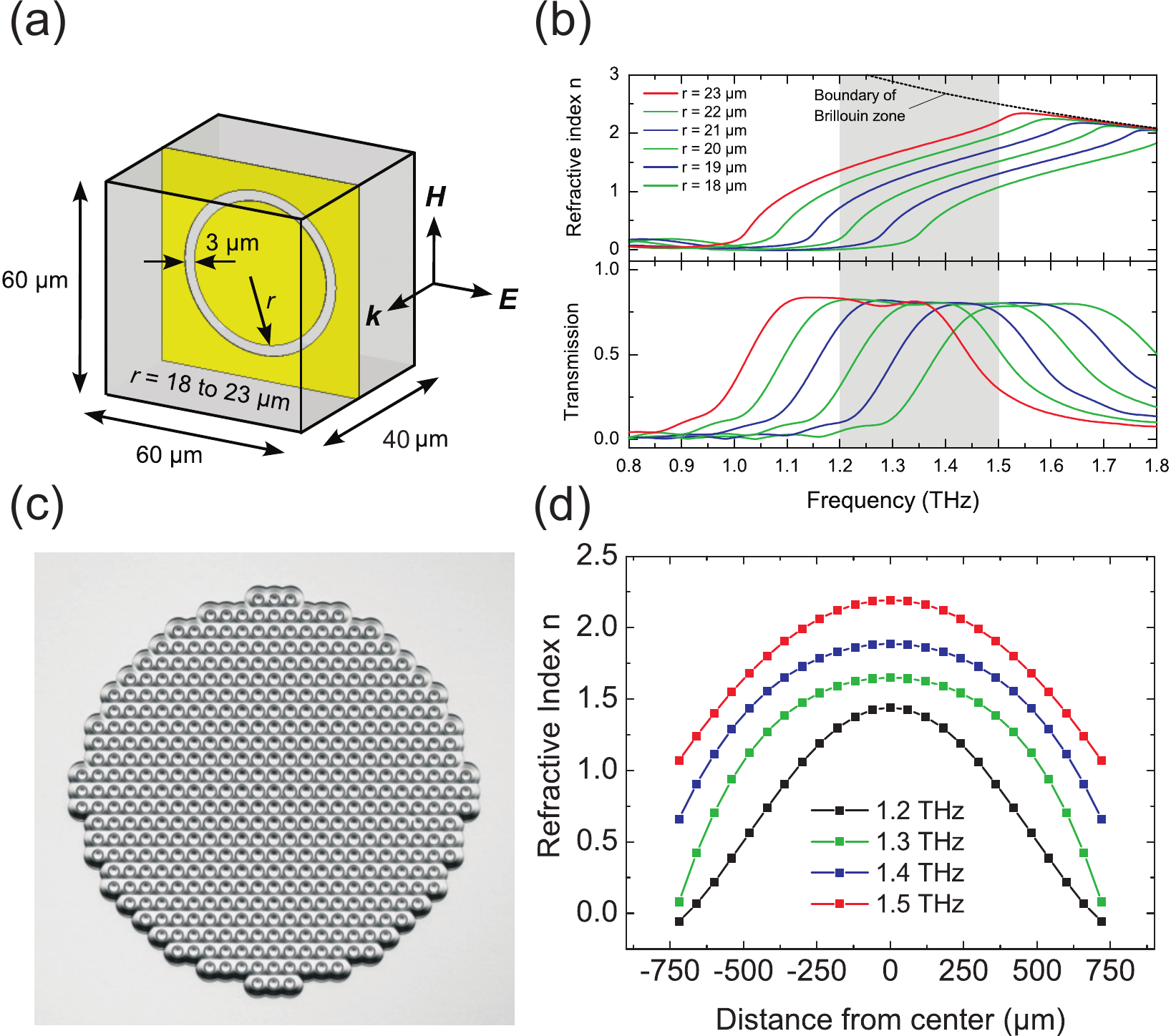}
\caption{\label{fig:Fig1} (a) Structural design of the unit cell. The underlying geometry was based on annular slots in a copper plane. (b) The refractive index of the metamaterial was changed by varying the inner radius of the annular slot between $r=18$ and 23\,\textmu m. This resulted in a total change of the refractive index from 0.08 to 1.65 at a frequency of 1.3\,THz. (c) Microscope picture of the 3-layer GRIN lens. (d) Refractive index in dependence of the radial distance from the center of the GRIN lens.}
\end{figure*}

Here we present the design, fabrication and the experimental testing of a three-layer metamaterial-based GRIN lens that allows one to focus THz radiation to a spot diameter smaller than the wavelength. We also compared the optical properties of the 3-layer GRIN lens to the focusing behavior of a 1-layer GRIN lens that is significantly easier to fabricate. We could show that a 1-layer lens already focuses THz radiation to a spot size in the order of the wavelength of the THz beam. The lenses offer broadband operation, are very thin and allow aberration-free optical imaging due to the avoidance of curved boundaries in the beam path. In contrast to most other materials, metamaterial lenses possess the potential to serve as adaptive optical components since they allow one in principle to tune the optical properties by optical or electronic means. In consequence, the presented GRIN lenses can be considered as an important step towards compact THz imaging systems with strongly improved spatial resolution.
\begin{figure*}
\centering
\includegraphics[scale = 0.4]{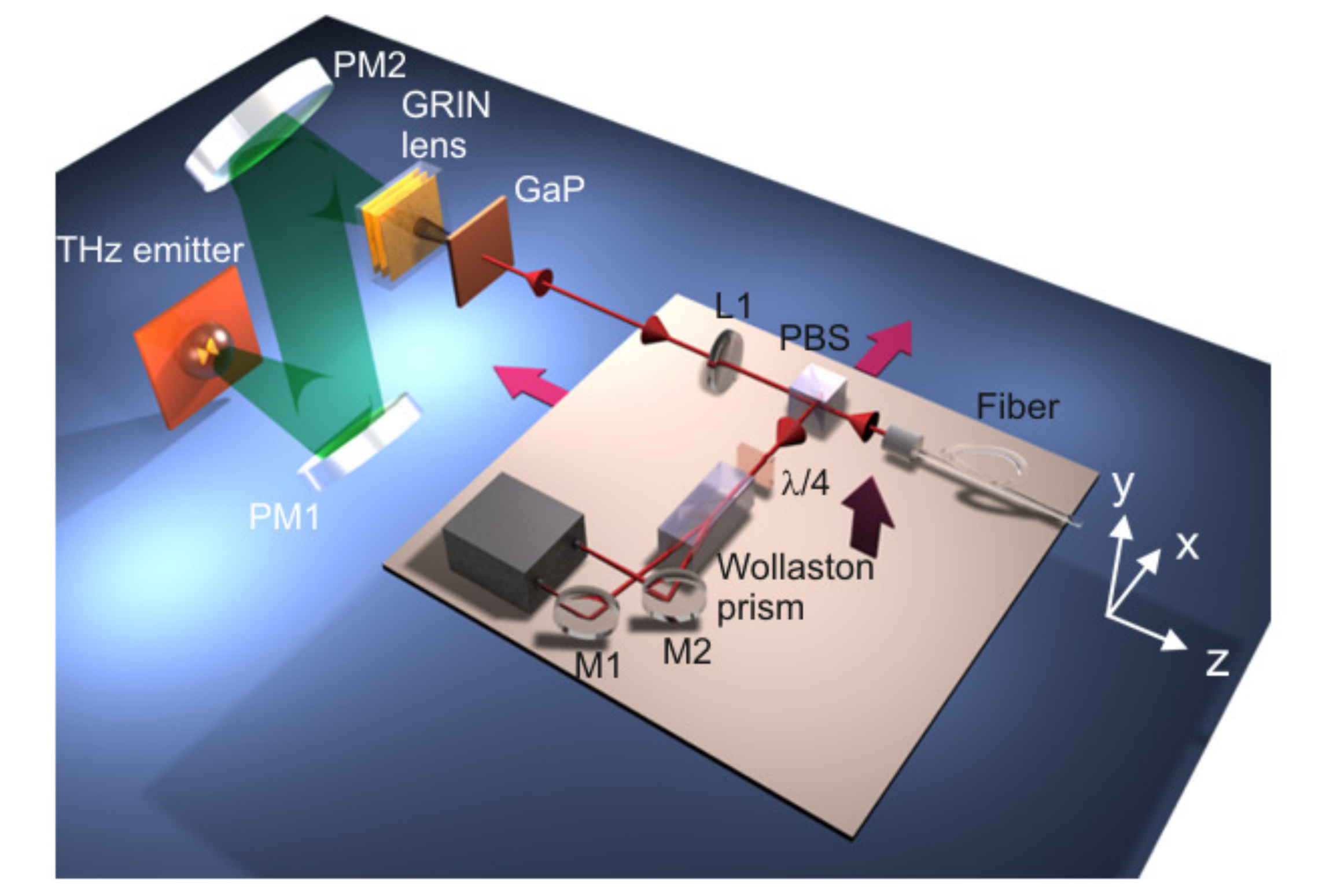}
\caption{\label{fig:Fig2} Schematic of the experimental setup for the beam profile measurement and the determination of the focus position of the GRIN lens. The THz electric field behind the lens is detected by exploiting the electro-optic effect in gallium phosphide (GaP). The optical probe beam is scanned over the THz beam to determine the spatial field distribution with a spatial resolution of 60\,\textmu m that is prescribed by the spot size of the optical beam. PM1,2: parabolic mirrors, L1: lens for the optical beam, M1,2: dielectric mirrors for the optical beam, $\lambda/4$: quarter-wave plate}
\end{figure*}

Fig.\ \ref{fig:Fig1}(a) illustrates the underlying structural design of the GRIN lens \cite{Paul:2010,Krug:1989}. The unit cell geometry was based on annular slots in a copper plane. Thereby, the refractive index of the unit cell could be strongly varied by changing the inner radius of the annular slot between $r=18$ and 23\,\textmu m at a constant slot width of 3\,\textmu m. The copper structure was fully embedded in a benzocyclobutene background matrix with a permittivity of $\epsilon=2.67$ and a loss tangent of $\delta=0.012$. The dimensions of the unit cell were $60^2\times 40$\,\textmu m$^3$. The copper layer had a thickness of 200\,nm and an electric conductivity of $\sigma=5.8\times 10^7$\,S/m. We retrieved the refractive index of the specific unit cell designs by use of three-dimensional full wave simulations with the software CST Microwave Studio. The numerical calculations showed that the investigated structure supports a refractive index gradient between the minimum and maximum values from 0.08 to 1.65 at a center frequency of 1.3\,THz (see Fig.\ \ref{fig:Fig1}(b)). For the determination of the extrema we restricted ourselves to the effective medium regime and took carefully the boundary to the Bragg regime into account since operation of the lens in the Bragg regime would lead to undesired spatial patterning of the focus and thus distort the optical quality of the focusing device. For the lens design we chose a radially symmetric refractive index gradient. For this purpose, we introduced a spatial variation of the refractive index by arraying unit cells of different inner slot radius such that the refractive index gradually decreased from the center of the GRIN lens. This is indicated in the microscope picture in Fig.\ \ref{fig:Fig1}(c) and Fig.\ \ref{fig:Fig1}(d). In more detail, the spatial dependence of the refractive index could be approximated by a parabolic index profile. Such a lens is conceived to focus incoming radiation since it is expected from the refractive index profile that an electromagnetic wave travels faster in the outer annular segments of the lens than in the inner annular segments. In this regard the refraction power correlates with the maximal achieved difference between the phase advance of the wave across the outer border and through the center of the GRIN lens. Thus, a thicker lens should result in a higher refraction power. For this reason, we compared the optical performance of a 1-layer and a 3-layer thick GRIN lens.
\begin{figure*}
\centering
\includegraphics[scale = 0.5]{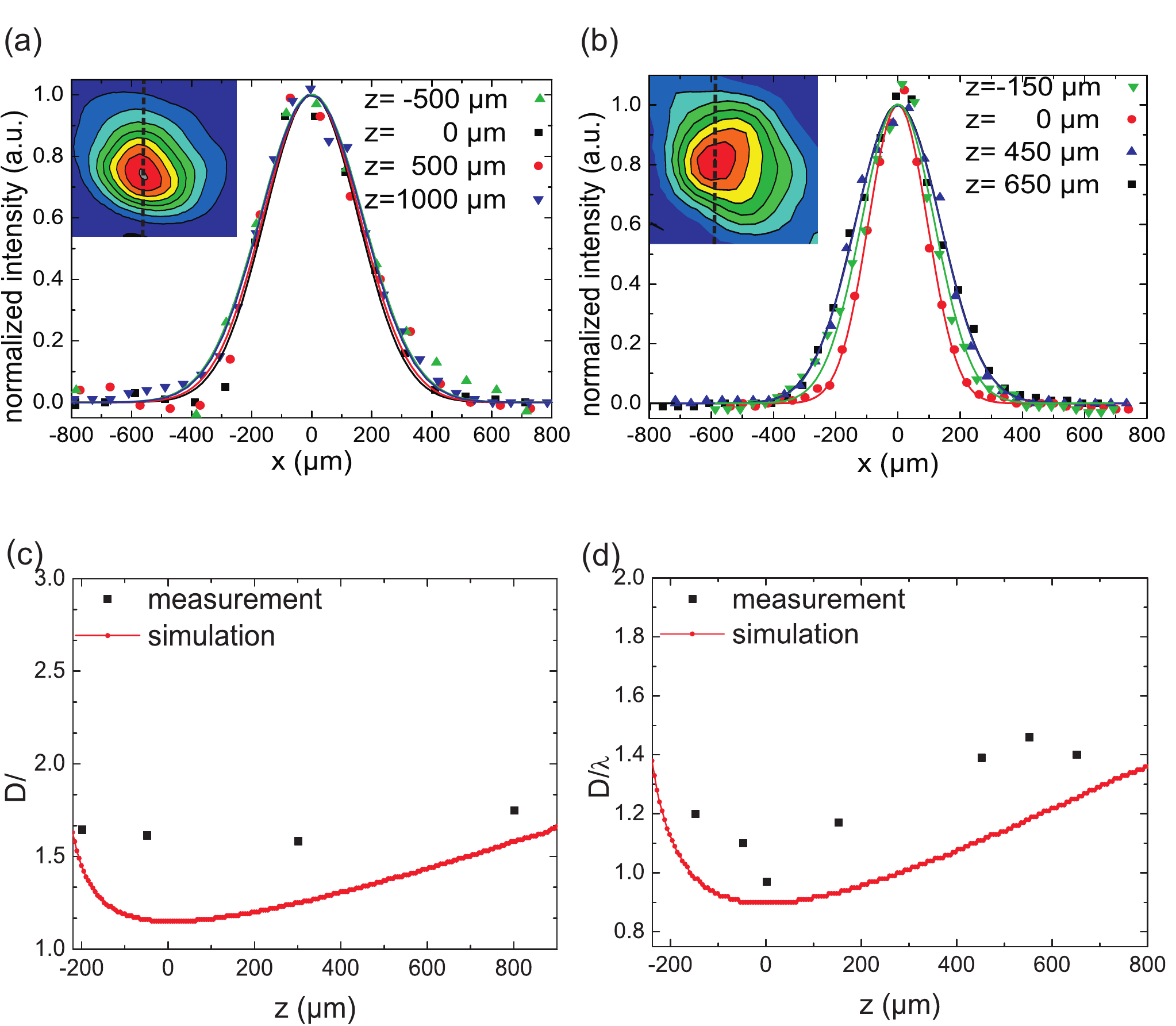}
\caption{\label{fig:Fig3} (a) 1D intensity profile of the THz beam for different z-positions in propagation direction of the THz wave for the 1-layer GRIN lens. The intensity profiles were obtained by extracting the intensity values along the 1D cross section line of the 2D transversal x-y-intensity profiles along the y-direction as shown in the inset. The z-position was measured relative to the focal plane. (b) Same as (a) for a 3-layer GRIN lens. (c) THz beam diameter $D$ as defined by the full width at half maximum (FWHM) of the intensity profile of the beam normalized to the wavelength $\lambda$ in dependence of the z-position in propagation direction of the THz wave for the 1-layer GRIN lens. For comparison, the theoretical data as obtained from 3D full wave simulations are plotted as solid lines. (d) Same as (c) for a 3-layer GRIN lens.}
\end{figure*}
\begin{figure*}
\centering
\includegraphics[scale = 0.7]{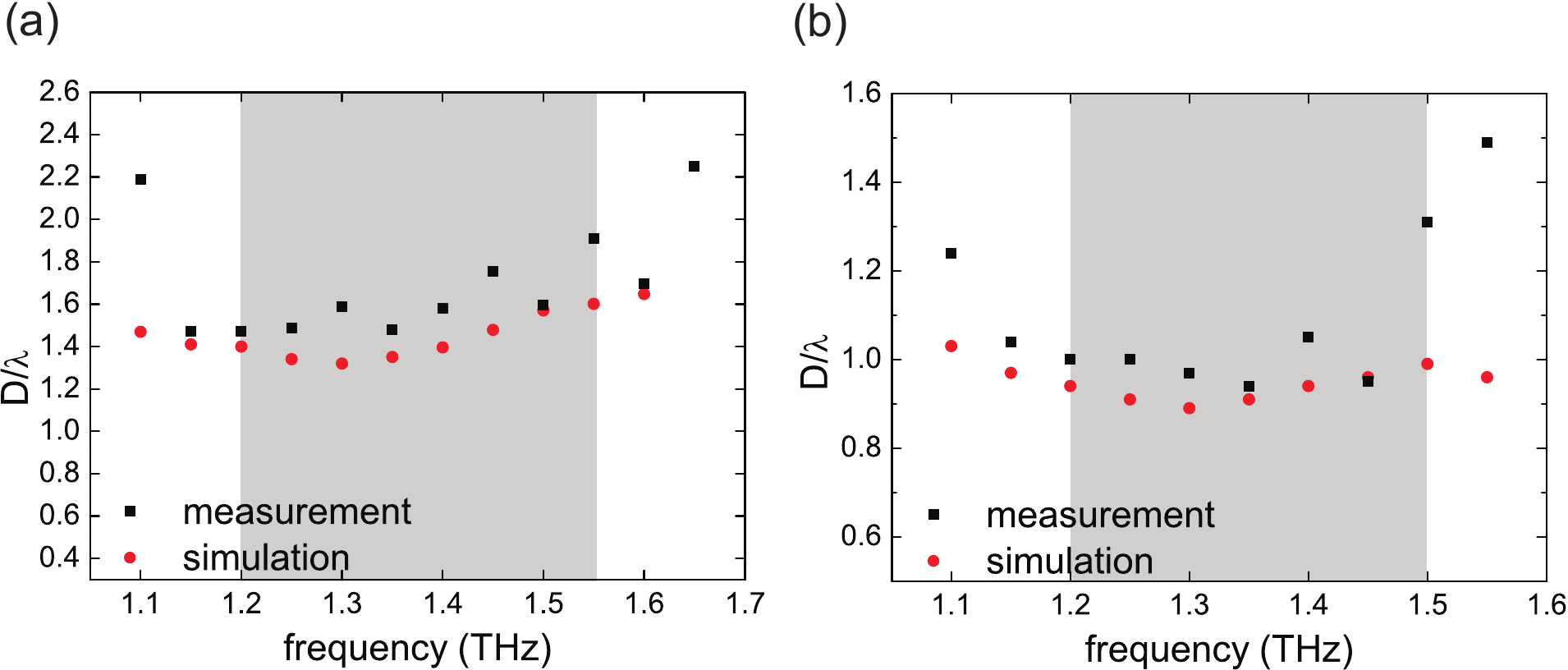}
\caption{\label{fig:Fig4} (a) Experimental and numerical values of $D/\lambda$ as defined by the ratio between the diameter $D$ of the THz beam and the center wavelength $\lambda$ of the THz radiation for the 1-layer GRIN lens. At frequencies higher than 1.5\,THz the lens no longer operated in the effective medium regime and the optical behavior was governed by scattering. Operation at frequencies higher than 1.5\,THz is therefore not recommended. (b) Same as (a) for the 3-layer lens. As envisioned by the design, the GRIN lens focused incident THz radiation to a focus diameter smaller than the wavelength. Over a frequency band between 1.2 and 1.5\,THz the ratio $D/\lambda$ was close to unity.}
\end{figure*}

We fabricated both the 1- and the 3-layer GRIN lens by UV lithography. That way we obtained a 1-layer GRIN lens with a thickness of 40\,\textmu m and a 3-layer lens of 120\,\textmu m thickness. We investigated the GRIN lenses with respect to their optical properties by mapping the spatial distribution of the complex electric THz fields. A schematic of the used experimental setup is shown in Fig.\ \ref{fig:Fig2}. We generated short THz pulses with a duration of 2\,ps and a spectral bandwidth from 0.1 to 2.0 THz by focusing ultrashort laser pulses with a duration of 15\,fs between the poles of a photoconductive switch. The emitted THz beam was polarized along the x-direction. In order to match the THz beam size to the aperture of the investigated GRIN lenses, we collimated the emitted THz pulses by an off-axis parabolic mirror PM1 and focused the THz beam to a spot diameter of about 1.5\,mm by a second parabolic mirror PM2. Hereby and in the following the spot diameter is defined by the full width at half maximum (FWHM) of the intensity distribution. We carefully aligned the GRIN lens with respect to the optical axis of the THz beam and positioned it in the beam waist of the THz radiation. By this method, we assured that the wave fronts of the incident THz radiation were plane at the entrance facet of the GRIN lens. Uttermost care was advised for the lateral alignment of the GRIN lens since the THz waves would have experienced an asymmetric refractive index gradient if slight lateral shifts of the center point of the GRIN lens had been introduced. Such a misalignment and the resulting asymmetric refractive index profile would cause highly undesirable beam shape distortions.

To determine the beam profile and the spot size of the focused THz beam we measured the spatial distribution of the complex electric field by use of electro-optic sampling in a reflection geometry \cite{Valk:2004,Knab:2009}. This configuration is illustrated in Fig.\ \ref{fig:Fig2} and the measurement technique is described in detail in the methods section of this letter.

In order to identify the position of the focal plane and the focus diameter of the 1- and 3-layer GRIN lens we measured the transversal x-y-intensity profile of the THz beam in dependence of the position z along the propagation direction. Examples of the obtained 2D intensity profiles are shown as insets of Figs.\ \ref{fig:Fig3}(a) and \ref{fig:Fig3}(b). From the 2D profiles we determined the 1D intensity distribution of the THz beam along the cross section line in y-direction through the center of the intensity distribution. This is illustrated in Fig.\ \ref{fig:Fig3}(a) for the 1-layer lens and in Fig.\ \ref{fig:Fig3}(b) for the 3-layer lens as measured at a center frequency of 1.3\,THz. Thereby, the position z was defined relative to the position of the corresponding focal plane. From the 1D intensity distributions we extracted the spot diameter $D$ in dependence of the z-position as illustrated in Fig.\ \ref{fig:Fig3}(c) and Fig.\ \ref{fig:Fig3}(d) for the 1- and 3-layer lens, respectively. For comparison, we also plotted the spot diameter that we obtained from numerical full wave calculations. As can be seen from Figs.\ \ref{fig:Fig3}(c) and \ref{fig:Fig3}(d), significant deviations of the theoretical spot diameter calculations from the experimental data are observed for the 1-layer lens whereas the agreement between the experimental and theoretical curves is very reasonable for the 3-layer lens. This was expected, since the 1-layer lens is more prone to surface effects that vary strongly with residual imperfections of the fabricated samples. In contrast, the 3-layer lens revealed bulk properties and the impact of surface effects on the electromagnetic properties of the lens was only minor.

As expected, both GRIN lens designs provoked a strong focusing of the THz radiation. By interpolation, we determined the focal plane of the 1-layer lens to be located 1.1\,mm behind the exit facet of the lens. From the measurements we determined a focal beam diameter of $D=320$\,\textmu m at 1.3\,THz which relates to $D/\lambda=1.45$ where $\lambda$ is the wavelength of the THz radiation. For the 3-layer lens, the focal plane was located 0.8 \,mm behind the lens exit surface. We measured a focus diameter of 220 \textmu m at 1.3\,THz which corresponds to a ratio $D/\lambda=0.96$ and thus indicates slight subwavelength focusing. A focus of subwavelength size was expected from numerical calculations for the 3-layer GRIN lens. The stronger focusing capabilities of the 3-layer lens in comparison with the 1-layer lens can be intuitively understood by considering the focus mechanism of a GRIN lens. The longer the propagation length through the GRIN lens is, the more the phase difference between the outer part and the inner part of the propagating wave accumulates. For this reason, we expect a larger curvature of the phase fronts of the beam which implicates a stronger focusing.

We also studied the spectral operation bandwidth of the lenses by measuring the diameter $D$ of the THz beam normalized to the wavelength $\lambda$ at the focal plane for different center frequencies (see Fig.\ \ref{fig:Fig4}). For comparison, we plotted the numerically calculated ratio $D/\lambda$ of the considered lenses in dependence of the center frequency. Equivalent to the foregoing discussion, the numerical and experimental results show better quantitative agreement for the 3-layer lens (Fig.\ \ref{fig:Fig4}(b)) than for the 1-layer lens (Fig.\ \ref{fig:Fig4}(a)). The measurements evidence that the GRIN lenses operated in a frequency range from 1.2 to 1.5\,THz. Within this range, the ratio $D/\lambda$ varied between 1.45 and 1.75 for the 1-layer lens and between 0.94 and 1.05 for the 3-layer lens. Therefore, we could achieve weak subwavelength focusing of THz radiation with a 3-layer GRIN lens over almost the complete frequency band for which the lens was designed. Note that the GRIN lenses operated in the Bragg regime for frequencies higher than 1.5\,THz (grey shaded region in Fig.\ \ref{fig:Fig4}). Since the beam shape was distorted by scattering in this frequency range the FWHM definition of the beam diameter was not longer valid. As mentioned before, lens operation in the scattering regime is not recommended.

In conclusion, we have demonstrated the design, fabrication and measurement of the optical properties of a 1-layer and a 3-layer GRIN lens and compared the optical performance of both lens types. We showed that a 1-layer GRIN lens already offers strong focusing capabilities and allows one to focus THz radiation to a spot diameter in the order of the wavelength. At a frequency of 1.3\,THz we obtained a focus diameter of $D=320$\,\textmu m. Even slight subwavelength focusing was expected and obtained for a 3-layer GRIN lens. We measured a focus diameter of $D=220$\,\textmu m at a frequency of 1.3\,THz which corresponds to a ratio between the focus diameter $D$ and the wavelength $\lambda$ of $D/\lambda=0.96$. Both lens types operated over a comparably large frequency range from 1.2 to 1.5 THz due to the non-resonant design of the metamaterial structure. The subwavelength focusing capabilities of the demonstrated metamaterial-based GRIN lenses in a broad frequency range can be considered as an important step towards the improvement of the spatial resolution of compact, industrial THz imaging systems.

\section*{Methods}
\subsection*{Measurement setup}
As an electro-optic material we employed a gallium phosphide (GaP) crystal with a thickness of 400\,\textmu m and an aperture size of 20$\times$20\,mm$^2$. The GaP crystal was cut in the (110)-direction such that the electro-optic effect was only sensitive to the x-polarization of the THz beam. The detection scheme was conceived as follows: Due to the electro-optic effect, the incident THz pulses induce an instantaneous change of the refractive index of the GaP crystal. This change is proportional to the momentary amplitude of the electric THz field. Since the electro-optic effect in GaP is anisotropic, it is possible to determine the amplitude of the electric THz field by measuring the change of the polarization state of ultrashort probe pulses that propagate through the crystal. For this purpose, we focused linearly polarized probe pulses to a focal spot size of 60\,\textmu m at the position of the GaP crystal. The wavelength of the probe pulses was 800\,nm and the pulse duration was 40\,fs. The surface of the GaP crystal facing the THz beam was coated by a highly reflective layer for a wavelength of 800\,nm while the opposite facet was equipped by an anti-reflection coating for the identical wavelength range. The optical coatings were designed in such a way that they did not affect the propagation properties of the THz beam. Consequently, the THz pulses could penetrate the GaP crystal without significant loss. On the other hand, we herewith ensured that the probe pulses could enter the GaP crystal with strongly reduced loss and were efficiently reflected at the other facet of the crystal where the THz pulses entered. During the propagation through the crystal the probe pulses changed their polarization state due to the optically induced change of the refractive index by the momentary THz field. Because of the anisotropy of the electro-optic effect, the difference in the induced phase delay of two orthogonally polarized field components of the probe pulses resulted in elliptically polarized probe pulses when the THz field was non-zero. Hereby, the orientation of the major and minor axes of the polarization ellipse was sensitive to the polarity and thus the phase of the momentary THz field. To enhance the sensitivity of the measurement we analyzed the polarization of the probe pulses by combination of a quarter wave plate, a wollaston prism and a balanced detector. Since we obtained the probe pulses from deviating a small fraction of the energy of the pump pulses used for the THz generation by a beam splitter, the probe pulses were inherently coherent with the generated THz pulses. Furthermore, the pulse length of 40\,fs of the probe pulses was a factor of 50 smaller than the pulse length of the THz pulses. This allowed us to sample the temporal shape of the electric field of the THz pulses by delaying the probe pulses with respect to the THz pulses. The coherence between the THz pulses and the probe pulses in combination with the phase sensitivity of the used detection scheme enabled us to measure both the amplitude and the phase of the electric field. Furthermore, since the optical probe beam was focused to a spot size of 60\,\textmu m the spatial resolution of the system was only limited by the spot size of the probe beam. Thus, we could measure the spatial field distribution of the THz beam with subwavelength resolution with respect to the wavelength of the THz radiation.

\hyphenation{fleisch-hauer}


\begin{thebibliography}{10}
\newcommand{\enquote}[1]{``#1''}

\bibitem{Dolling:2007}
G.~Dolling, M.~Wegener, C.~M. Soukoulis, and S.~Linden, Opt. Express
  \textbf{15}, 11536 (2007).

\bibitem{Ergin:2010}
T.~Ergin, N.~Stenger, P.~Brenner, J.~B. Pendry, and M.~Wegener, Science
  \textbf{328}, 337 (2010).

\bibitem{Freymann:2010}
G.~von Freymann, A.~Ledermann, M.~Thiel, I.~Staude, S.~Essig, K.~Busch, and
  M.~Wegener, Advanced Functional Materials \textbf{20}, 1038 (2010).

\bibitem{Xiao:2010}
S.~Xiao, V.~P. Drachev, A.~V. Kildishev, X.~Ni, U.~K. Chettiar, H.-K. Yuan, and
  V.~M. Shalaev, Nature \textbf{466}, 735 (2010).

\bibitem{Lagarkov:2010}
A.~N. Lagarkov, V.~N. Kisel, and A.~K. Sarychev, J. Opt. Soc. Am. B
  \textbf{27}, 648 (2010).

\bibitem{Fang:2010}
A.~Fang, T.~Koschny, and C.~M. Soukoulis, Journal of Optics \textbf{12}, 024013
  (2010).

\bibitem{Zheludev:2008}
N.~I. Zheludev, S.~L. Prosvirnin, N.~Papasimakis, and V.~A. Fedotov, Nature
  Photonics \textbf{2}, 351 (2008).

\bibitem{Plumridge:2008}
J.~Plumridge, E.~Clarke, R.~Murray, and C.~Phillips, Solid State Communications
  \textbf{146}, 406 (2008).

\bibitem{Kaestel:2005}
J.~Kästel and M.~Fleischhauer, Phys. Rev. A \textbf{71}, 011804 (2005).

\bibitem{Liu:2009}
N.~Liu, L.~Langguth, T.~Weiss, J.~Kästel, M.~Fleischhauer, T.~Pfau, and
  H.~Giessen, Nature Materials \textbf{8}, 758 (2009).

\bibitem{Rayspan:2010}
{Rayspan Corporation}, Website (2010). Available online at
  \url{http://www.rayspan.com/index.htm}.

\bibitem{Hoshina:2009}
H.~Hoshina, Y.~Sasaki, A.~Hayashi, C.~Otani, and K.~Kawase, Appl. Spectrosc.
  \textbf{63}, 81 (2009).

\bibitem{Rutz:2006}
F.~Rutz, M.~Koch, S.~Khare, M.~Moneke, H.~Richter, and U.~Ewert, International
  Journal of Infrared and Millimeter Waves \textbf{27}, 547 (2006).

\bibitem{Herrmann:2002}
M.~Herrmann, M.~Tani, K.~Sakai, and R.~Fukasawa, Journal of Applied Physics
  \textbf{91}, 1247 (2002).

\bibitem{Taday:2004}
P.~F. Taday, Phil. Trans. R. Soc. Lond. A \textbf{362}, 351 (2004).

\bibitem{Mittleman:1996}
D.~M. Mittleman, R.~H. Jacobsen, and M.~C. Nuss, IEEE Journal of Selected
  Topics in Quantum Electronics \textbf{2}, 679 (1996).

\bibitem{Tonouchi:2007}
M.~Tonouchi, Nature Photonics \textbf{1}, 97 (2007).

\bibitem{May:2007}
T.~May, S.~Anders, V.~Zakosarenko, M.~Starkloff, H.-G. Meyer, G.~Thorwirth, and
  E.~Kreysa, Proc. SPIE \textbf{6549}, 65490D (2007).

\bibitem{Yamamoto:2004}
K.~Yamamoto, M.~Yamaguchi, F.~Miyamaru, M.~Tani, M.~Hangyo, T.~Ikeda,
  A.~Matsushita, K.~Koide, M.~Tatsuno, and Y.~Minami, Japanese Journal of
  Applied Physics \textbf{43}, L414 (2004).

\bibitem{Weg:2009}
C.~am~Weg, W.~von Spiegel, R.~Henneberger, R.~Zimmermann, T.~Loeffler, and
  H.~Roskos, Journal of Infrared, Millimeter and Terahertz Waves \textbf{30},
  1281 (2009).

\bibitem{Kemp:2003}
M.~C. Kemp, P.~F. Taday, B.~E. Cole, J.~A. Cluff, A.~J. Fitzgerald, and W.~R.
  Tribe, Proc. SPIE \textbf{5070}, 44 (2003).

\bibitem{Chen:2006}
H.-T. Chen, W.~J. Padilla, J.~M.~O. Zide, A.~C. Gossard, A.~J. Taylor, and
  R.~D. Averitt, Nature \textbf{444}, 597 (2006).

\bibitem{Chen:2008}
H.-T. Chen, J.~F. O’Hara, A.~K. Azad, A.~J. Taylor, R.~D. Averitt, D.~B.
  Shrekenhamer, and W.~J. Padilla, Nature Photonics \textbf{2}, 295 (2008).

\bibitem{Chen:2009}
H.-T. Chen, W.~J. Padilla, M.~J. Cich, A.~K. Azad, R.~D. Averitt, and A.~J.
  Taylor, Nature Photonics \textbf{3}, 148 (2009).

\bibitem{Paul:2009}
O.~Paul, C.~Imhof, B.~Lägel, S.~Wolff, J.~Heinrich, S.~Höfling, A.~Forchel,
  R.~Zengerle, R.~Beigang, and M.~Rahm, Opt. Express \textbf{17}, 819 (2009).

\bibitem{Landy:2008}
N.~I. Landy, S.~Sajuyigbe, J.~J. Mock, D.~R. Smith, and W.~J. Padilla, Phys.
  Rev. Lett. \textbf{100}, 207402 (2008).

\bibitem{Tao:2008}
H.~Tao, N.~I. Landy, C.~M. Bingham, X.~Zhang, R.~D. Averitt, and W.~J. Padilla,
  Opt. Express \textbf{16}, 7181 (2008).

\bibitem{Paul:2009a}
O.~Paul, R.~Beigang, and M.~Rahm, Opt. Express \textbf{17}, 18590 (2009).

\bibitem{Weis:2009}
P.~Weis, O.~Paul, C.~Imhof, R.~Beigang, and M.~Rahm, Applied Physics Letters
  \textbf{95}, 171104 (2009).

\bibitem{Paul:2010}
O.~Paul, B.~Reinhard, B.~Krolla, R.~Beigang, and M.~Rahm, Applied Physics
  Letters \textbf{96}, 241110 (2010).

\bibitem{Krug:1989}
P.~A. Krug, D.~H. Dawes, R.~C. McPhedran, W.~Wright, J.~C. Macfarlane, and
  L.~B. Whitbourn, Opt. Lett. \textbf{14}, 931 (1989).

\bibitem{Valk:2004}
N.~C.~J. van~der Valk, T.~Wenckebach, and P.~C.~M. Planken, J. Opt. Soc. Am. B
  \textbf{21}, 622 (2004).

\bibitem{Knab:2009}
J.~R. Knab, A.~J.~L. Adam, M.~Nagel, E.~Shaner, M.~A. Seo, D.~S. Kim, and
  P.~C.~M. Planken, Opt. Express \textbf{17}, 15072 (2009).

\end{thebibliography}
\end{document}